\documentclass[useAMS,usenatbib]{mn2e}

\usepackage{epsfig,graphicx,amsmath,amssymb}
\usepackage{times}
\usepackage{natbib}

\def\prd{Phys. Rev. D}
\def\aap{A\&A}
\def\apj{ApJ}

\def\apjl{ApJ}
\def\mnras{MNRAS}

\def\aj{AJ}

\def\physrep{Phys. Rep.}

\def\apjs{ApJS}

\title[Cosmic reionization in a dynamic quintessence cosmology]
{Cosmic reionization in dynamic quintessence cosmology}
\author[D.~Crociani et al.]
{D. Crociani$^1$\thanks{email: daniela.crociani2@studio.unibo.it},  M. Viel$^{2,3}$, 
L. Moscardini$^{1,4}$, M. Bartelmann$^5$, M. Meneghetti$^6$\\
$^1$ Dipartimento di Astronomia, Universit\`a di Bologna, via Ranzani  
1, I-40127 Bologna, Italy \\
$^2$ INAF-Osservatorio Astronomico di Trieste, via G.B. Tiepolo 11,
I-34131 Trieste, Italy \\
$^3$ INFN/National Institute for Nuclear Physics, Sezione di Trieste, 
via Valerio 2, I-34127 Trieste, Italy \\
$^4$ INFN/National Institute for Nuclear Physics, Sezione di Bologna, 
viale Berti Pichat 6/2, I-40127 Bologna, Italy \\
$^5$ Zentrum f\"ur Astronomie, ITA, Universit\"at Heidelberg, Albert-\"Uberle-Str. 2,
D-69120 Heidelberg, Germany\\
$^6$ INAF-Osservatorio Astronomico di Bologna, via Ranzani 1, I-40127 Bologna, Italy}

\begin{document}

\date{Accepted ???. Received ???; in original August 2007}

\pagerange{\pageref{firstpage}--\pageref{lastpage}} \pubyear{2007}

\maketitle

\label{firstpage}

\begin{abstract}
In this paper we investigate the effects that a dynamic dark energy
component dominant in the universe at late epochs has on reionization.
We follow the evolution of HII regions with the analytic approach of
\citet{furlanetto2005} in two different universes for which we assume
the \citet{peebles2003} and \citet{brax2000} quintessence models and
we compare our results to the $\Lambda$CDM scenario.  We show that,
for a fixed ionization efficiency, at the same cosmological epoch the
morphology of bubbles is dominated by high-mass objects and the
characteristic size of the ionized regions is slightly smaller than in
the $\Lambda$CDM model, especially at the latest stages of
reionization, due to the higher recombination efficiency.  As a
consequence, the bubbles' `epoch of overlap' happens earlier than in
$\Lambda$CDM. Finally, we show how the different evolution of the HII
regions affects the transmission of the high-$z$ QSO spectra, reducing
the Lyman flux absorption at small optical depths.
\end{abstract}



\section{Introduction} \label{sect:intro}

Reionization of the intergalactic medium (IGM) is a crucial epoch for
the history of the universe, when the neutral gas produced at
recombination is ionized by the Ultra-Violet (UV) radiation emitted by
the first luminous sources.  After this stage, the IGM contains a
small amount of neutral gas, responsible for the absorptions in the
spectra of background objects.  Reionization is still a poorly
understood process because of the unknown nature of the ionizing
sources and of the complex physical mechanisms producing the radiation
emission.  However, as suggested by recent observations, it turns out
to be better described as a spatially inhomogeneous and not
instantaneous phase \citep[see, e.g.,][and references
therein]{Ciardi2005}.

While the Lyman-$\alpha$ transmitted flux in the high-$z$ quasar (QSO)
spectra found by the Sloan Digital Sky Survey suggests that the end of
reionization is at $z\sim 6$
\citep{fan2001,becker2001,white2003b,fan2006}, the last results from
the Cosmic Microwave Background (CMB) polarization provide a Thompson
optical depth of $\tau=0.09\pm 0.03$ that requires the completion of
the reionization process at $z\sim 10$ \citep{spergel2007}.  On the
other hand, the IGM temperature measurements at $z<4$ \citep{hui2003}
show that the reionization epoch occurs at $6<z<10$, while the lack of
evolution in the Lyman-$\alpha$ galaxy luminosity function at $5.7\la
z\la 6.5$ suggests that probably half of the IGM is ionized at $z\sim
6.5$ \citep{malhotra2004}.  More recent estimates, based on high-$z$
HIRES QSOs, seem instead to argue against sudden changes in the IGM
properties due to late reionization at $z\sim 6$
\citep{Becker:2006qj}: it is worth stressing, however, that the use of
QSO near zones to probe the IGM ionization fraction could still be
problematic \citep[see, e.g.,][]{Bolton2007}.

Future observations, such as Lyman-$\alpha$ galaxy surveys
\citep{kashicawa2006}, measurements of the CMB polarization and of the
Sunyaev-Zeldovich effect that will be obtained by the Planck
satellite, and, more importantly, neutral hydrogen 21 cm observations
through new generation telescopes (LOFAR, SKA), must provide further
information to constrain the reionization scenario.  In recent years,
several analytic, semi-analytic and numerical models 
\citep[see, e.g.][]{wyithe2003,barkana2004, haiman2003,madau2004,choudhury2006,
gnedin2000,ciardi2003b,Wyithe2006,iliev2007} have been developed in
order to describe how the first sources of UV-radiation impact on the
IGM. The basic assumption of these approaches is to model the
relations between the HII regions, the ionizing sources that allows to
describe the morphology of bubbles, and the galactic physics, such as
gas cooling, star formation, radiative feedback and radiative
transfer. Despite some controversial results, the reionization process
fits reasonably well in a ``standard'' $\Lambda$CDM cosmology, i.e. in
a model where cold dark matter (CDM) and baryon density fluctuations
grow in a flat universe dominated at late epochs by a dark energy
component, consisting in cosmological constant. The latter is
characterized by a constant equation of state $p=w\rho c^{2}$, with
$w=-1$.

At present, the best probe of dark energy is provided by SNe of type
Ia. However, even including the most recent high$-z$ observations
\citep[see, e.g.,][]{riess2007}, it is still not possible to obtain a high-precision
determination of the dark energy equation. Tight constraints could
only be put if strong and unjustified priors on its evolution are
assumed.  While observations suggest that $w\sim -1$ at late epochs,
the time evolution of the dark energy component is basically not
constrained. The so-called quintessence models, where $w$ is varying
in time, are not excluded.  In the CDM cosmologies with dynamic dark
energy, the main consequence of having $w>-1$ at high $z$ is the
earlier growth of the matter fluctuations. This determines a higher
number density of haloes in the quintessence universe than in the
standard cosmology at a fixed epoch \citep[see, e.g.][]{Maio2006}.
Thereby, the presence of a dark energy component might affect the
reionization process, requiring a different ionization efficiency to
fully ionize the IGM at a given redshift.

In this work, we use an analytic approach to investigate how
reionization proceeds in quintessence cosmologies.  In doing this, we
consider two different cosmological models for which we assume that
the redshift dependence of the equation of state parameter $w(z)$
follows the self-interacting potentials proposed by
\citet{peebles2003} and \citet{brax2000}.  The predicted scenario
obtained by ``painting'' the evolution of the HII regions through the
\cite[][hereafter F05]{furlanetto2005} analytic model, is compared to
that expected for a $\Lambda$CDM universe.

The paper is organized as follows. In Section \ref{sect:qd} we briefly
outline the quintessence cosmologies considered here.  In Section
\ref{sect:model} we review the main features of the
F05 model. Section \ref{sect:res} contains the results on the
evolution of the ionized bubbles and their properties. The final
Section \ref{sect:conclu} summarises our main conclusions.


\section{Probing the dynamic quintessence cosmology} \label{sect:qd}

The main aim of this paper is to investigate the process of cosmic
reionization in quintessence cosmologies and compare to the predicted
scenario for a standard flat $\Lambda$CDM universe.  Thus, the
$\Lambda$CDM cosmology will be our reference case, for which we assume
that the contributions to the present density parameter from
cosmological constant, matter, and baryons are
$\Omega_{0\Lambda}=0.7$, $\Omega_{\rm 0m}=0.3$ and $\Omega_{\rm
0b}=0.046$, respectively; the Hubble constant is $H_{0}=70$ km/s/Mpc
(i.e. $h=0.7$ in units of 100 km/s/Mpc).  We also fix the
normalization of the power spectrum of the matter fluctuations
according to $\sigma_{8}=0.9$ and the spectral index is $n=1$.  These
parameters are in agreement with the WMAP first-year data
\citep{spergel2003}. We recall that the more recent analysis of the
WMAP three-year data \citep{spergel2007} suggests slightly different
values (in particular a lower $\sigma_{8}$ and a smaller $\Omega_{\rm
0m}$). Our parameter choice, which is done to allow a direct
comparison with the similar analysis made by F05,
can have some small quantitative effect on some of the results, but
cannot alter the general conclusions of our analysis, which is aimed at 
discussing the expected differences between
models where the dark energy component is provided by a cosmological
constant or by a dynamic quintessence.
 
\begin{figure}
\begin{center}
\includegraphics[width=8.5cm, height=8.5cm]{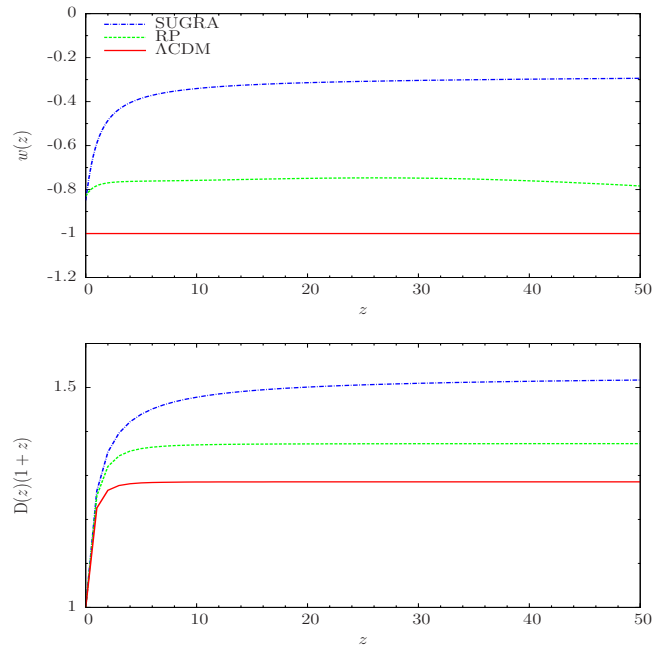}
\end{center}
\caption{
Redshift evolution of the cosmic equation-of-state parameter $w$ (top
panel) and of the growth factor, given in term of $D(z)(1+z)$ and
normalized to its value at the present time (bottom panel).  Different
lines refer to $\Lambda$CDM model (solid line), RP model (dashed line)
and SUGRA model (dotted-dashed line).}
\label{fig:1b}
\end{figure}

Thus, the dark energy models we consider are cosmological scenarios in
which the dynamic dark energy component is characterized by a
self-interacting scalar field $\Phi$, evolving under the effects of
the potential $V(\Phi)$. Here we summarize the main features of these
models \citep[see][for more details]{peebles2003}.

The potential has to satisfy the Klein-Gordon equation:
\begin{equation}\label{eq:1b}
\ddot{\Phi}+3H(z)\dot{\Phi}+\frac{\partial V}{\partial\Phi}=0\ ,
\end{equation}
where $H(z)$ represents the redshift evolution of the Hubble constant
given by the usual Friedmann equation.

\begin{figure*}
\includegraphics[width=15cm, height=5cm]{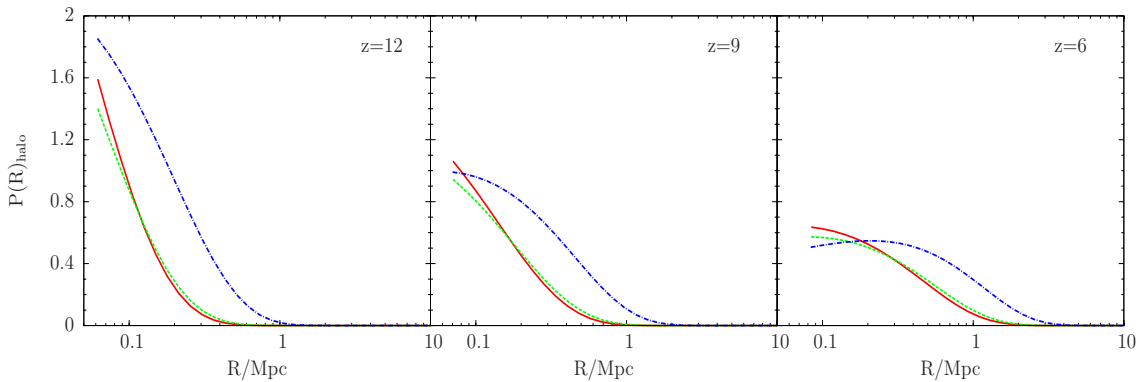} 
\caption{
The halo distribution at different epochs: redshifts $z=12$, $z=9$,
and $z=6$ are shown in the panels, from left to right. For each
cosmological model, the halo mass function is computed consistently to
the PS74 formalism and written in terms of the halo radius.  Different
curves refer to $\Lambda$CDM model (solid line), RP model (dashed
line) and SUGRA model (dotted-dashed line).}
\label{fig:2b}
\end{figure*}

The corresponding $z$-evolution of the quintessence parameter $w$ is
provided by the equation of state
\begin{equation}\label{eq:2b}
w\equiv \frac{p}{\rho
c^{2}}=\frac{\frac{\dot{\Phi}^{2}}
{2}-V(\Phi)}{\frac{\dot{\Phi}^{2}}{2}+V(\Phi)}\ ,
\end{equation}
such that $w\rightarrow-1$ if $\dot{\Phi}^{2}\ll V(\Phi)$. Many
analytic expressions have been proposed for the potential
$V(\Phi)$. They describe the present dark energy amount simply by
setting their amplitude on the initial conditions and determine
different redshift evolutions for $w$. In our analysis, we use the
potentials proposed by \citet{peebles2003} (RP hereafter),
\begin{equation}\label{eq:3b}
V(\Phi)=\frac{k}{\Phi^{\alpha}}\ ,
\end{equation}
and by \citet{brax2000}
\begin{equation}\label{eq:4b}
V(\Phi)=k\Phi^{-\alpha}\exp\Big(\frac{\Phi^{2}}{2}\Big)\ ,
\end{equation}
as suggested by supergravity theory (SUGRA hereafter); $k$ has
dimension of mass raised to the power $(\alpha+4)$.  The values for
$k$ and $\alpha$ are fixed by assuming a flat universe with a dark
energy contribution to the present density parameter $\Omega_{\rm
0de}=0.7$, and $w_{0}\equiv w(z=0)=-0.85$ and $-0.83$ for SUGRA and RP
potentials, respectively.  These choices for $w_0$, even if only
marginally consistent with the observational constraints \citep[see,
e.g.][]{riess2007}, have been made with the purpose of emphasizing
differences with the $\Lambda$CDM model.  The remaining parameters
have been fixed to those of the $\Lambda$CDM model: $h=0.7$,
$\sigma_8=0.9$ and $n=1$.  Note that the quintessence models here
considered do not violate the constraints recently obtained from the
WMAP three-year data \citep{spergel2007} for the electron scattering
optical depth,
\begin{equation}\label{eq:5b}
\tau_{e}=\int_{0}^{z_{r}}n_{e}(z)\sigma_{T}\frac{dL(z)}{dz}dz\ ,
\end{equation}
computed at the reionization epoch, assumed to be $z_{r}=6$.  In the
previous equation, $\sigma_{T}$ represents the Thompson scattering
cross section, and $n_{e}(z)$ and $L(z)$ are the electron density and
the comoving distance, respectively. In more detail, we estimate for
the $\Lambda$CDM, RP and SUGRA models $\tau_{e}=0.132, 0.130$ and
$0.125$, which agree (at 1.5$\sigma$ level) with the WMAP optical
depth.  Among all the possible dark energy models, we concentrate
on these two since they have been accurately investigated by other
authors under the theoretical point of view and their impact on
observational quantities have been extensively addressed \citep[see,
e.g.,][]{dolag2004,meneghetti2005}.  Furthermore, deviations from the
$\Lambda$CDM behaviour are larger at high redshifts, in an interesting
regime for the reionization analysis made here. This is evident by
looking at the time evolution of the equation-of-state parameter $w$,
shown in the upper panel of Fig.~\ref{fig:1b}. Though the RP and SUGRA
display similar values for $w$ today, strong differences appear at
high redshifts. Parametrizing the evolution of $w$ in terms of the
expansion factor $a$ as $w(a)=w_{0}+w_{a}(1-a)$, we find that the RP
and SUGRA models can be fitted by $w_{a}\sim 0.08$ and $0.55$
respectively. These values are still consistent with the present
observational constraints. For example, \citet{Liddle2006} combining
data from CMB, SNIa and baryonic acoustic oscillations found
$w_{a}=0.0\pm 0.8$. Of course our results will depend on the
considered dark energy scenarios: models having parameters more
similar to (different from) the $\Lambda$CDM cosmology (for which
$w_0=-1$, $w_a=0$) would show smaller (larger) effects on the
observables discussed here. These high-redshift differences affect
the initial phases of the reionisation process. In particular, we
expect that at the redshifts of interest $(z\ga 6)$ the RP model
behaves as an `intermediate case' between SUGRA and $\Lambda$CDM.
Note that in quintessence cosmologies, the growth factor $D(z)$ is
larger than in the standard cosmology, as shown in the lower panel of
Fig.~\ref{fig:1b}, where the quantity $D(z)(1+z)$ has been normalized
to its value at the present time. Thus, the main effects of having $w
> -1$ at high redshifts are an earlier formation of the structures and
a higher abundance of haloes than in the $\Lambda$CDM cosmology at a
fixed epoch \citep[see the discussion in][]{Maio2006}. As an
illustrative example, in Fig.~\ref{fig:2b} we show the distribution of
the dark matter halo sizes as predicted by the \citet{press1974}
theory (PS74) for the $\Lambda$CDM (solid line), RP (dashed line) and
SUGRA (dotted-dashed line) cosmologies at three different redshifts.
Indeed, at the same cosmological epoch, the halo distribution is
dominated by larger objects in the quintessence cases, in particular
in the SUGRA model. This can strongly affect the ionization process of
the universe. Indeed numerical simulations
\citep{ciardi2003,sokasian2003,sokasian2004} showed that reionization
is an `inside-out' phenomenon, i.e. it begins in overdense regions and
expands in underdense regions: this property predominantly
originates from the fast increase of the abundance of ionising
sources. However, we remark that a slower (or vanishing) change of
the source population should cause a rapid escape of the ionising
photons towards the external underdense regions: in this case
reionization would be an outside-in process.


\section{An analytic approach to cosmic reionization} 
\label{sect:model}

To investigate how reionization occurs we use the analytic approach
proposed by F05. In this section we review the main features of the
model: for further details we refer to the original paper and to
\citep[][hereafter F04]{furlanetto2004a}.

\subsection{Evolution of bubbles without recombination}

The evolution of the ionized bubbles is determined by the hierarchic
growth of matter fluctuations. This can be described by making use of
the extended PS74 formalism \citep{lacey1993}.  A warning has to
be kept in mind about the use of the Press \& Schechter mass function,
that is proven to not work well for rare haloes at high redshifts. In
particular, numerical simulations \citep[see, e.g.,][]{iliev2006b,reed2007,
lukic2007} show that it underestimates their abundance by a significant
factor, with not negligible effects on the bubble distribution.
However, we recall that the extended formalism has not be developed
for mass functions different from PS74: for this reason we have to
rely on it, even if this can affect some of the the following
results.

In the PS74 scenario, at a fixed cosmological epoch, an ionized bubble
grows around a galaxy of mass $m_{\rm gal} \ge m_{\rm min}(z)$, where
$m_{\rm min}(z)$ represents the virial mass corresponding to
$T=10^{4}$K, the temperature at which hydrogen cooling becomes
efficient.  The mass associated to the ionized region is
$m_{HII}=\zeta m_{\rm gal}$, where $\zeta$ represents the ionization
efficiency of the galaxy (here assumed to be constant), that depends
on the star formation rate, on the escape fraction of photons and on
the number of HII recombinations.  Since each region is thought to be
isolated, it must contain enough collapsed mass to fully ionize the
inner gas. Thus $f_{\rm coll}(\delta,m) \ge \zeta^{-1}$, where $f_{\rm
coll}$ is the collapsed volume fraction of a region of mass m $\ge
m_{\rm min}$ with an inner overdensity $\delta$.  In the F04
formalism, this leads to the following condition for the overdensity
inside a bubble of a given mass $m$ (in Lagrangian space):
\begin{equation}
\label{eq:1c}
\delta_{m} \ge \delta_{x}(m,z) \equiv 
\delta_{c}(z)-\sqrt{2}K(\zeta)[\sigma^{2}_{\rm min}-
\sigma^{2}(m)]^{1/2}\ ,
\end{equation}
where $K(\zeta) \equiv {\rm erf}^{-1}(1-\zeta^{-1})$, $\sigma^{2}(m)$
is the variance of density fluctuations smoothed on the scale $m$ and
$\sigma^{2}_{\rm min} \equiv \sigma^{2}(m_{\rm min})$. As shown in
F04, $\delta_{x}$ represents the ionization
threshold for the density fluctuations in the Lagrangian space and it
is assumed to be a linear barrier with respect to $\sigma^{2}(m)$:
$\delta_{x}(m,z) \sim B(m,z)=B_{0}(z)+B_{1}(z)\sigma^{2}(m)$. Hence it
is possible to obtain an analytic expression for the distribution of
the bubbles with mass in the range $m\pm {\rm d}m/2$:
\begin{equation}
\label{eq:2c}
n(m,z)=\sqrt{\frac{2}{\pi}}\frac{\bar{\rho}}{m^{2}}\Big\vert\frac{{\rm
d}\ln\sigma}{{\rm d}\ln m}\Big\vert\frac{B_{0}(z)}
{\sigma(m)}\exp\Bigg[-\frac{B^{2}(m,z)}{2\sigma^{2}(m)}\Bigg]\ ,
\end{equation}
where $\bar{\rho}$ is the mean comoving matter density of the
universe. In a similar way, adopting the \citet{lacey1993} formalism,
we can write the merger rate of the HII regions as:
\begin{eqnarray}\label{eq:3c}
\frac{{\rm d}^{2} p(m_{1},m_{T},t)}{{\rm d} m_{2}{\rm d}
t}&=&\sqrt{\frac{2}{\pi}}\frac{1}{t} \Big\vert\frac{{\rm d}\ln B}{{\rm
d}\ln t}\Big\vert\Big\vert\frac{{\rm d}\ln\sigma_{T}}{{\rm d}\ln
m_{T}}\Big\vert\times\nonumber\\
&&\Big(\frac{1}{m_{T}}\Big)\frac{B(m_{T},z)}
{\sigma_{T}(1-\sigma^{2}_{T}/\sigma^{2}_{1})^{3/2}}\times\nonumber\\
&&\exp\Bigg[-\frac{B_{0}^{2}(z)}{2}
\Bigg(\frac{1}{\sigma^{2}_{T}}-\frac{1}{\sigma^{2}_{1}}\Bigg)\Bigg]\ ,
\end{eqnarray}
where ${\rm d}^{2} p(m_{1},m_{T},t)/{\rm d} m_{2}{\rm d} t$ is the
probability per unit time that a given halo of mass $m_{1}$ merges
with a halo of mass $m_{2}=m_{T}-m_{1}$.  From equation (\ref{eq:3c}),
it is possible to define the merger kernel
\begin{equation}\label{eq:14c}
Q(m_{1},m_{2},t)\equiv \frac{1}{n(m_{2},t)}
\frac{{\rm d}^{2} p(m_{1},m_{T},t)}{{\rm d} m_{2}{\rm d}t}\ ,
\end{equation}
that represents the rate at which each region of mass $m_{1}$ merges
with a region of mass $m_{2}$. Since this quantity suffers from some
limitations because the asymmetry in its arguments becomes important
for large masses, then the use of $Q_{sym}(m_{1},m_{2})\equiv
1/2[Q(m_{1},m_{2})+Q(m_{2},m_{1})]$ is preferred for estimating the
merger rate of the bubbles. This allows us to define the fractional
volume accretion for a bubble of mass $m_{1}$ that merges with a mass
$m_{1}$:
\begin{eqnarray}\label{eq:15c}
V(m_{1})^{-1}\frac{{\rm d}V}{{\rm d}z}&\equiv&
\frac{V(m_{2})}{V(m_{1})}m_{2}n(m_{2},z)\nonumber\\
&&\times Q_{sym}(m_{1},m_{2},t)\Big\vert\frac{{\rm d}t}{{\rm
d}z}\Big\vert.
\end{eqnarray}
Finally, we recall that the global ionized fraction can be calculated
as $\bar{x}_{i}=\zeta f_{\rm coll,g}(z)$, where $f_{\rm coll,g}$ is
the global collapsed volume fraction.

\subsection{The recombination-limit effects}

Up to now, the recombination limit has been neglected. As a bubble
grows, the photons propagate more deeply into the neutral IGM, and
both the clumpiness and the recombination rate of the ionized gas
increase. The IGM distribution and its ionization state can be
described using the analytic model of \citet{miralda2000}
(MHR00). Analysing numerical simulations at $z < 4$, they found an
analytic expression for the volume-weighted distribution of the gas
density:
\begin{equation}\label{eq:4c}
P_{V}(\Delta)=A_{0}\Delta^{-\beta}\exp\Bigg[-\frac{(\Delta^{-2/3}-C_{0})^{2}
}{2(2\delta_{0}/3)^{2}}\Bigg]\ ,
\end{equation}
where $\Delta\equiv \rho/\bar{\rho}$, $\delta_{0}$ is the r.m.s. of
density fluctuations smoothed on the Jeans mass at fixed $z$, so
$\delta_{0}\propto (1+z)^{-1}$; $A_{0}$ and $C_{0}$ represent
normalization constants and $\beta$ can be set equal to 2.5 as
predicted for isothermal spheres.  The ionization state of the IGM is
determined by a density threshold $\Delta_{i}$ such that the gas with
$\Delta < \Delta_{i}$ is totally ionized and the gas with $\Delta >
\Delta_{i}$ is neutral. Under this assumption, the recombination rate
can be written as
\begin{equation}\label{eq:5c}
A(\Delta_{i})=A_{u}\int^{\Delta_{i}}_{0}P_{V}(\Delta)\Delta^{2}{\rm
d}\Delta \equiv A_{u}C\ ,
\end{equation}
where $C$ represents the clumping factor and $A_{u}$ is the
recombination rate per hydrogen atom in gas at the mean density. In
the F05 model, $A_{u}$ is assumed consistently with the A-case of the
MHR00 model \citep[see also][]{miralda2003}: $A_{u}\propto
\alpha_{A}(T)$, where $\alpha_{A}(10^{4} K)=4\times 10^{-13}$
cm$^{3}$s$^{-1}$. The MHR00 model also provides a relationship between
the mean free path $\lambda_{i}$ of the photons and the ionized
fraction of gas $F_{V}(\Delta_{i})$, namely:
\begin{equation}\label{eq:6c}
\lambda_{i}=\lambda_{0}[1-F_{V}(\Delta_{i})]^{-2/3}\ ,
\end{equation}
where $\lambda_{0}$ is a normalization constant such that
$\lambda_{0}H(z)=60$ km s$^{-1}$ at $z < 4$.

In the following, we assume that the mean free path derived in the
MHR00 model could be used also in the quintessence models, since the
properties of this function should reflect the gas properties at the
Jeans scale, which is much smaller than the scales we are interested
in. However, this approximation deserves further investigation with
suitable hydrodynamical simulations.

To consider the recombination process it is necessary to relate the
recombination rate to the smoothed matter overdensity. In doing this,
it must be remarked that the main effect of an inhomogeneous gas
distribution is an increasing gas clumpiness and subsequently an
increasing HII recombination rate. As a consequence,
$A_{u}\propto(1+\delta)$.  When a bubble grows the ionizing photons
are able to reach its edge, then the threshold must satisfy the
condition $\lambda_{i}(\Delta_{i}) \ge R$ that sets
$\Delta_{i}$. However, at the same time, the inner high gas clumpiness
causes an increase of the recombination rate and the photons can be
absorbed inside the bubble before reaching the edge. Then, for a
growing region, the ionization rate has to be larger than the
recombination rate at every time:
\begin{equation}\label{eq:17c}
\zeta\frac{{\rm d}f_{\rm coll}(\delta,R)}{{\rm d}t}>A_{u}C(R)(1+\delta)\ ,
\end{equation} 
where $C(R)$ is computed as in equation (\ref{eq:5c}) for
$R=\lambda_{i}$. The recombination barrier is obtained by searching
for the minimun $\delta$ in the Lagrangian space that satisfies
equation (\ref{eq:17c}) at each given mass.

The recombination process affects the bubbles geometry. When the
ionizing photons are totally absorbed by the inner recombination, the
HII region stops growing and reaches a maximum size $R_{\rm max}$
that, in a $\Lambda$CDM universe, slowly increases with decreasing
redshift, as shown by F05.

In the excursion-set formalism, the recombination limit has a deep
impact on the distribution of the bubbles. Assuming that the
recombination barrier is a vertical line crossing $\delta_{x}$ at
$R_{\rm max}$, the trajectories such that $\delta(R_{\rm max})<
B(R_{\rm max},z)$ will be incorporated into HII regions with $m <
m_{\rm max}$ and the mass function reads:
\begin{equation}\label{eq:7c}
n_{\rm rec}(m,z)=\int^{B(R_{\rm max})}_{-\infty}p(\delta\vert R_{\rm max})
n(m,z\vert\delta,m_{\rm max},z){\rm d}\delta\ .
\end{equation}
In the previous equation $p(\delta\vert R_{\rm max})$ represents the
probability distribution at the scale $R_{\rm max}$ for a Gaussian
density field
\begin{equation}\label{eq:8c}
p(\delta\vert R_{\rm max})=
\frac{1}{\sqrt{2\pi}\sigma_{\rm max}}\exp\Bigg(-\frac{\delta^{2}}{2\sigma^{2}_{\rm 
max}}\Bigg)\ ,
\end{equation}
and $n(m,z\vert\delta,m_{\rm max},z)$ is the conditional mass function
for a random walk that begins at $(\delta,\sigma^{2}_{\rm max})$
\begin{eqnarray}\label{eq:9c}
n(m,z\vert\delta,m_{\rm max},z)&=&\sqrt{\frac{2}{\pi}}\frac{\bar{\rho}}{m^{2}}\Big\vert
\frac{{\rm d}\ln\sigma}{{\rm d}\ln m}\Big\vert\times\nonumber\\
&&\frac{\sigma^{2}[B(m_{\rm
max},z)-\delta]}{(\sigma^{2}-\sigma^{2}_{\rm
max})^{3/2}}\times\nonumber\\
&&\exp\Bigg\{-\frac{[B(m,z)-\delta]^{2}}{2(\sigma^{2}-\sigma^{2}_{\rm
max})}\Bigg\} \ ,
\end{eqnarray}
where $\sigma_{\rm max}\equiv \sigma(R_{\rm max})$. Since every
trajectory lying above the ionization barrier at $R_{\rm max}$
belongs to a bubble with $R=R_{\rm max}$, the distribution of such
Str\"omgren regions can be obtained from equation (\ref{eq:8c}):
\begin{equation}\label{eq:10c}
N_{\rm rec}=\frac{\bar{\rho}}{2m_{\rm max}}{\rm
erfc}\Bigg[\frac{B(R_{\rm max},z)}{\sqrt{2}\sigma_{\rm max}}\Bigg]\ .
\end{equation}

After the saturation, bubbles can grow only by merging. For a single
point of the IGM, reionization ends when it is incorporated in a
recombination-limited region, since the ionizing background slightly
increases after merging of bubbles. Thus `overlap' is a local
phenomenon. The volume fraction of the IGM in bubbles with $R > R_{\rm
max}$ is provided by the PS74 formalism using the ionization barrier
and results to be:
\begin{equation}\label{eq:11c}
x_{\rm rec}=\int^{+\infty}_{m_{\rm max}}n(m,z)V(m){\rm d}m\ ,
\end{equation}
where $V(m)$ is the volume of the ionized region. 

\subsection{The Lyman-$\alpha$ flux transmission}

The morphology of bubbles affects the absorption of the Lyman-$\alpha$
flux emitted from high-redshift sources. Indeed, large ionized regions
allow the transmission of the Lyman-$\alpha$ forest because the extent
of neutral regions in the IGM is large enough to reduce the
Lyman-$\alpha$ `damping wing' absorption, as pointed out by
\citet{furlanetto2004c}. As shown by F05, since the
recombination process affects the late stages of reionization, the way
how the bubbles saturate can be constrained through the Ly-flux
transmission. Therefore the observed transmitted flux from high-$z$
galaxies can constrain the predicted evolution of HII regions.

Including the recombination limit in the F05 model allows to write the
probability of having an optical depth smaller than $\tau_{i}$ for the
$i$-th transition, determined by the distribution of the IGM and by
the morphology of the ionized bubbles, as follows:
\begin{equation}\label{eq:12c}
P(<\tau_{i})=\int^{+\infty}_{m_{HII\rm min}}n(m,z)\frac{m}{\bar{\rho}}{\rm d}m
\int^{\Delta_{\rm max}}_{0}P_{V}(\Delta){\rm d}\Delta\ ,
\end{equation}    
where $m_{HII\rm min}$ is the minimum mass of the HII regions and
$\Delta_{\rm max}$ is the maximum density for which
$\tau<\tau_{i}$. The analytic expression for the inner overdensity of
each bubble is obtained by assuming the equation of state for the
polytropic gas $T=T_{0}\Delta^{\gamma}$, with $T_{0}=10^{4}$K and the
ionization equilibrium inside each bubble. Under these assumptions,
the neutral hydrogen fraction can be written as a function of the
matter overdensity $\Delta$:
\begin{equation}\label{eq:18c}
x_{HI}=\frac{\chi_{e}\bar{n}_{H}\alpha(T)}{\Gamma}\Delta\ ,
\end{equation}
where $\chi_{e}$ is the correction for the singly-ionized helium and
$\Gamma$ is the ionizing rate per hydrogen atom. It mainly depends on
the total photons' emissivity $\epsilon_{T}$ and on the mean free path
$\lambda_{i}$ as
\begin{equation}\label{19c}
\Gamma\propto\lambda_{i}\epsilon_{T}\Big(\frac{\eta}{3+\eta}\Big)\ .
\end{equation}
At the end of reionization $\epsilon_{T}\propto \zeta {\rm d}f_{\rm
coll,g}/{\rm d}t$, $\eta=3/2$ if a starburst spectrum is assumed and
$\lambda_{i}$ is set to the minimum value between the bubble radius
and $R_{\rm max}$. Finally, $P_{V}(\Delta)$ is thought to be
independent of the bubble morphology, that could be a good approximation
at the end of reionization although we need high-resolution
simulations to test it. Hence, the relation between the local
overdensity and the IGM optical depth is:
\begin{eqnarray}\label{eq:13c}
\Delta(\tau_{i})&=&\Bigg\{170\frac{\eta}{3+\eta}\frac{\alpha_{A}(10^{4}{\rm
K})}{\alpha_{A}(T_{0})}h(z)\Bigg(\frac{\lambda_{i}}{\rm
Mpc}\Bigg)\times\nonumber\\ &&\zeta\Big\vert\frac{{\rm
d}f_{\rm coll}}{{\rm d}z}\Big\vert\Bigg(\frac{\tau_{i}}
{\tau_{GP,i}}\Bigg)\Bigg\}^{1/(2-0.7\gamma)}\ ,
\end{eqnarray}
where $\tau_{GP,i}$ is the Gunn \& Peterson optical depth for the
$i$-th transition. Note that in the equation above the value of the
Hubble parameter $h(z)$ is taken consistently from the different
cosmological models. Finally, the probability for the inhomogeneous
IGM to have a given optical depth $\tau$ is obtained by substituting
$\Delta_{\rm max}$ in equation (\ref{eq:12c}).

A warning has to be kept in mind regarding the application of our model
to derive the Lyman fluxes.  Some of the simplifying assumptions
present in the description of the recombination process, like the abrupt 
change of the IGM ionization state as a function of its density
\citep[see, e.g., discussion in][]{miralda2000}. We expect, however,
that a more realistic treatment would change our predictions in the
same way for the different cosmological models.  For this reason, we
prefer to present our results in terms of ratios between the fluxes
derived for the quintessence models and those predicted for the
$\Lambda$CDM model.


\section{Results and discussion}  \label{sect:res}

In this section we present the main results of the application of the
previous model to cosmological scenarios including a dynamic
quintessence (see Section \ref{sect:qd}). Notice that F04 and F05
apply their model to the `concordance' $\Lambda$CDM model only.

First of all, we compute the minimum collapsed mass at each
cosmological epoch from the mass-temperature relation proposed by
\citet{barkana2001}:
\begin{eqnarray}\label{eq:1d}
T_{\rm
vir}&=&1.98\times10^{4}\Big(\frac{\mu}{0.6}\Big)
\Big(\frac{M}{10^{8}h^{-1}M_{\odot}}\Big)^{2/3}\times\nonumber\\
&&\Big[\frac{\Omega_{\rm m}}{\Omega_{\rm m}^{z}}\frac{\Delta_{c}}
{18\pi^{2}}\Big]^{1/3}\Big(\frac{1+z}{10}\Big)
{\rm K}\ ,
\end{eqnarray}
where $T_{\rm vir}$ is the virial temperature of a halo having mass
$M$ and $\mu$ is the mean molecular weight of its inner gas.  In
order to take into account the fact that the IGM inside a HII region is
not totally ionized because of the ionization equilibrium assumption,
here we prefer to set $\mu$ to the mean of the values discussed in
\citet{barkana2001}. Anyway, we checked that fixing the molecular weight
to the value corresponding to a fully ionized IGM ($\mu=0.6$) does
not significantly affect the model predictions on the observables
discussed at the end of this section. For example, using $\mu=0.6$ changes
the Ly-$\alpha$ flux transmission by $\sim 1\%$
only, irrespectively of the considered cosmological model.

\begin{figure}
\begin{center}
\includegraphics[angle=0, height=4cm, width=7cm]{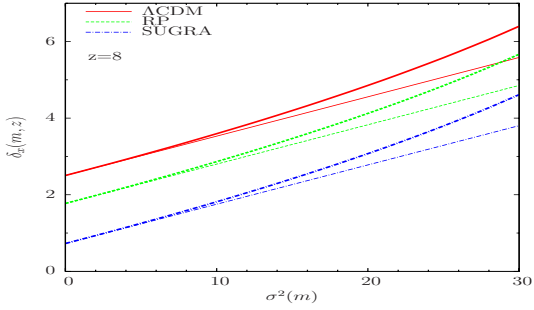}
\caption{The ionization barrier 
is here shown at $z=8$, assuming an ionization efficiency $\zeta=6$.
For each cosmological model, the thick line represents $\delta_{x}$ as
defined in equation (\ref{eq:1c}), while the thin line is its linear
approximation. The curves refer to the $\Lambda$CDM model (solid
line), RP model (dashed line) and SUGRA model (dotted-dashed line).}
\label{fig:1d}
\end{center}
\end{figure}

The values for $\Delta_c$ and for the redshift evolution of $\Omega_{\rm m}(z)$
are computed consistently for the different cosmologies. The presence
of a dynamic quintessence component affects the hierarchic evolution
of the ionized regions. Indeed, the earlier growth of matter
perturbations causes, at a fixed epoch, the formation of larger
ionised regions in the quintessence cosmology.  Another important
consequence is that, even if the assumption of a linear ionization
barrier is still correct, this is lower than in the $\Lambda$CDM
universe. This is evident in Fig.~\ref{fig:1d}, where we show the
behaviour of the ionization barrier in the three different
cosmological models, computed at $z=8$, assuming $\zeta=6$.

\begin{figure*}
\includegraphics[angle=0, height=5cm, width=15cm]{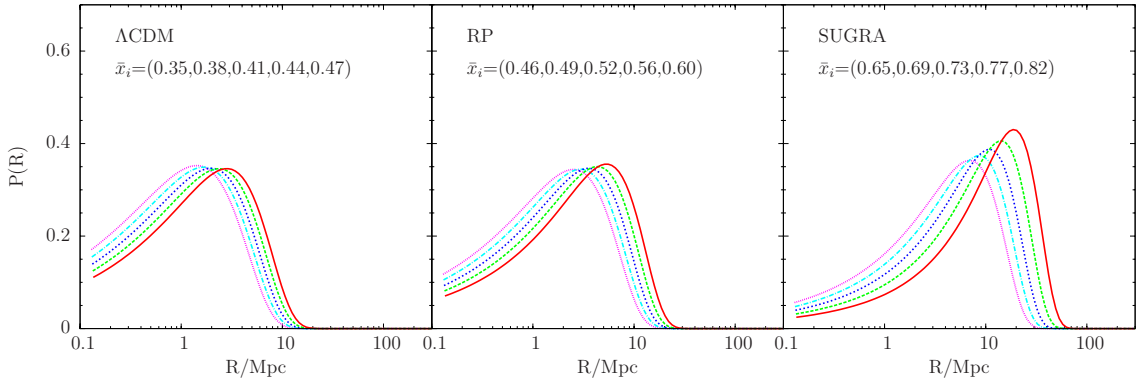}
\caption{
The morphology of the HII regions. The probability distribution of the
bubble sizes is computed here neglecting recombination.  In each panel
(corresponding to different cosmological models), the curves refer to
different epochs: $z=8.8$ (dotted line), 8.6 (dotted-dashed line), 8.4
(short-dashed line), 8.2 (long-dashed line) and 8 (solid line).  The
corresponding values for the ionized volume fraction $\bar{x}_{i}$,
computed assuming a ionization efficiency $\zeta=6$, are also
reported.}
\label{fig:2d}
\end{figure*}

This implies a higher probability that matter fluctuations go beyond
the ionization threshold. Consequently the density of bubbles
increases, resulting in a different morphology of HII regions in the
three cosmologies considered, as shown in Fig.~\ref{fig:2d}.  The
earlier growth of the matter fluctuations in the quintessence models
causes a faster evolution of the mass function with respect to the
$\Lambda$CDM universe and a remarkable increase of the density of the
largest regions. As a consequence, at the same cosmological epoch, we
obtain a higher ionized fraction $\bar{x}_{i}$ in the quintessence
cases. As an illustrative example, we observe that at $z=8.8$,
reionization is at its initial phases in the `standard' universe
($\bar{x}_{i}=0.35$) while for the SUGRA cosmology this epoch
corresponds to its late stages, $\bar{x}_{i}=0.65$.  This is almost a
factor $\sim$ 2 larger than in the `standard' case. Similarly, at
$z=8$ reionization is at the final stage in the SUGRA universe,
$\bar{x}_{i}=0.82$, while $\bar{x}_{i}=0.47$ for the `concordance'
model.  We recall that the model we adopt, being based on the
extended PS74 formalism, can account only in a very approximate way,
most often in the linear limit, for the source clustering,
which is expected to have important effects on the morphology of the
HII regions.  Since the clustering amplitude depends on the source
abundances which are different in the models considered here, our
results could be affected by this bias.  Work is in progress to
properly address this problem by improving our modelling with suitable
numerical simulations.

\begin{figure*}
\begin{center}
\includegraphics[angle=0, height=5cm, width=15cm]{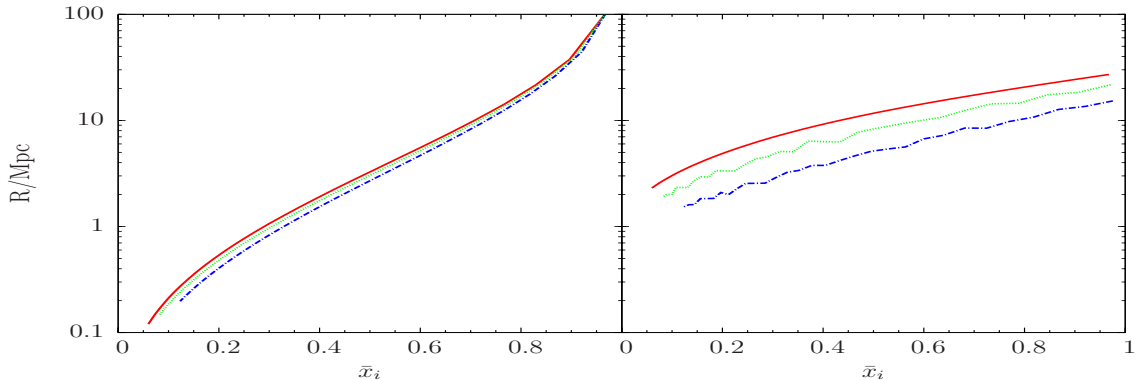}
\caption{
The bubble morphology.  The evolution of the bubble radius as
a function of $\bar{x}_{i}$ is shown here neglecting
recombination (left panel) and in the recombination limit (right panel).
Solid, dotted and dotted-dashed lines are for $\Lambda$CDM, RP and SUGRA models, respectively. 
A complete reionization at $z=6$ is assumed here. }
\label{fig:3d}
\end{center}
\end{figure*}

\begin{figure*}
\includegraphics[angle=0, height=5cm, width=15cm]{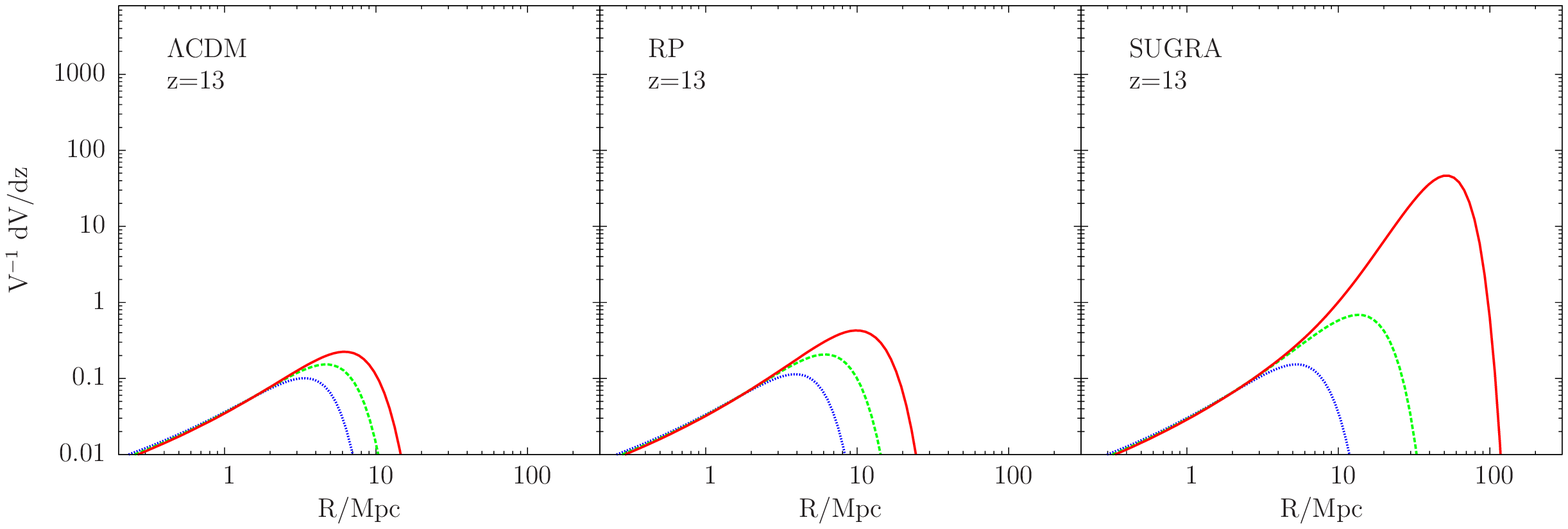}
\caption{The hierarchic growth of bubbles. 
In each panel (corresponding to different cosmological models), we
show the fractional volume accretion rate of an HII region having mass
$M_{1}=10^{14}M_{\odot}$ that merges with a mass corresponding to a
given radius $R$. The results, computed at $z=13$, are obtained
assuming different ionization efficiencies: $\zeta=10$ (dotted line),
$\zeta=20$ (dashed line) and $\zeta=30$ (solid line).}
\label{fig:4d}
\end{figure*}

In Fig.~\ref{fig:3d} we show the different characteristic sizes,
$R_{\rm char}$, of the ionized regions, as obtained by fixing $\zeta$
such that reionization ends at $z=6$ (note that $\zeta$ is not the
same for the three models).  Neglecting the recombination limit, the
effects of the quintessence are obvious at the early stages of
reionization, since $R_{\rm char}$ (representing the radius for which
the bubbles distribution is maximum) is larger in the standard
universe than in RP and SUGRA models. For example, at $x_{i}\simeq
0.2$ $R_{\rm char}=0.4, 0.6$ Mpc for $\Lambda$CDM and SUGRA,
respectively. But the difference becomes increasingly smaller as
reionization proceeds. This is caused by the different saturation
regime of the IGM in the quintessence cosmologies. Ionized regions are
smaller in the RP and SUGRA models at the beginning of reionization
and their sizes evolve faster than in the $\Lambda$CDM universe due to
the presence of large neutral voids around them, reaching the
characteristic radius of the standard model in the final stages of
reionization.

Since the HII regions grow only by merging after the recombination
limit, it is interesting to investigate how the dynamic dark energy
component affects the bubbles' merger rates. As an example,
Fig.~\ref{fig:4d} shows the merger probability of a region with mass
$M=10^{14} M_{\odot}$ at $z=13$ for the $\Lambda$CDM, RP and SUGRA
universes. Similarly to the $\Lambda$CDM case, also for the
quintessence cosmologies the evolution of the bubbles is dominated by
merging events between large systems, in particular in the late stages
of the reionization process. The main difference is given by the size
of the involved regions. For the RP and SUGRA models the merger
probability is higher with bubbles that are even one order of
magnitude larger than in the standard universe.

\begin{figure*}
\includegraphics[angle=0,height=9cm, width=8.3cm]{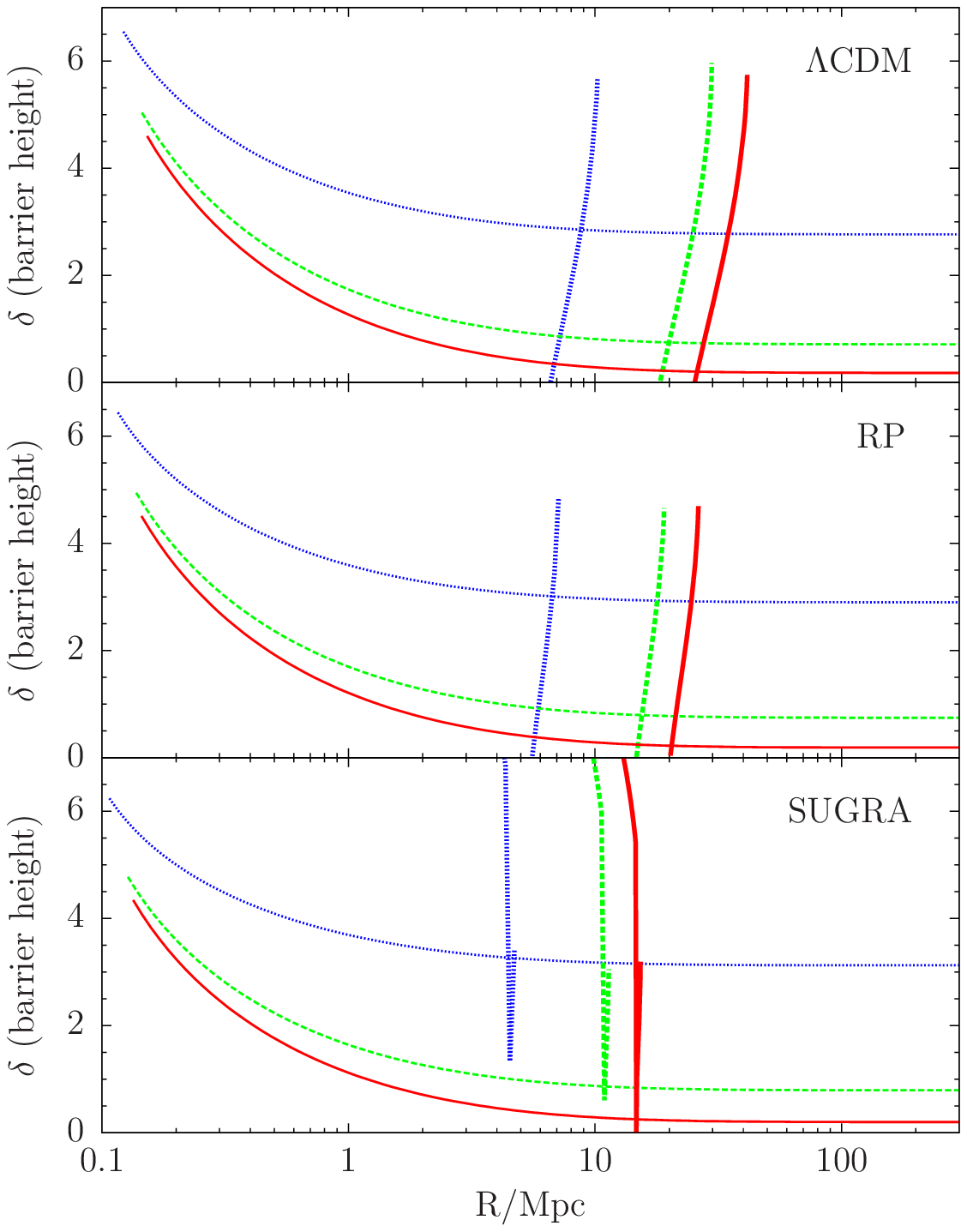}
\caption{
The recombination limit.  In each panel (corresponding to different
cosmological models), we show the ionization barrier, computed as in
F04 (thin curves), and the recombination barrier, computed as in F05
(thick curves).  The results are shown at three different stages of
reionization: $\bar{x}_{i}=0.49$ (dotted lines), $\bar{x}_{i}=0.82$
(dashed lines) and $\bar{x}_{i}=0.95$ (solid lines) at $z=6$.}
\label{fig:5d}
\end{figure*}
\begin{figure*}
\includegraphics[angle=0,height=5cm,width=15cm]{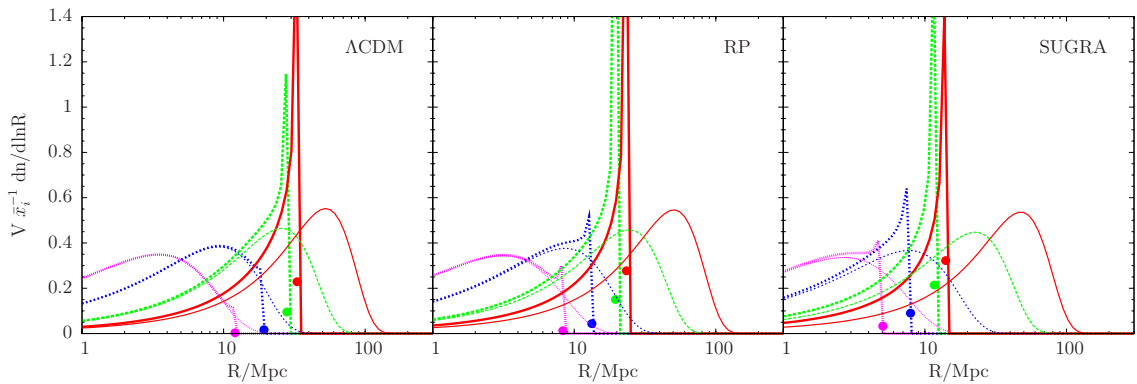}
\caption{
The bubble size distribution considering the recombination process.
The mass distributions show that in the recombination limit (thick
curves) bubbles pile up at $R=R_{\rm max}$, while neglecting
recombination (thin curves) they can reach sizes larger than $R_{\rm
max}$.  In each panel (corresponding to different cosmological
models), the different curves refer to the HII regions distribution at
different stages of reionization: $\bar{x}_{i}=0.51$ (dotted line),
$\bar{x}_{i}=0.68$ (short-dashed line), $\bar{x}_{i}= 0.84$
(long-dashed line), $\bar{x}_{i}=0.92$ (solid line) assumed for $z=8$.
The filled points represent the volume fraction in bubbles with
$R=R_{\rm max}$ in the recombination limit.}
\label{fig:6d}
\end{figure*}

\begin{figure*}
\includegraphics[angle=0,height=5cm,width=15cm]{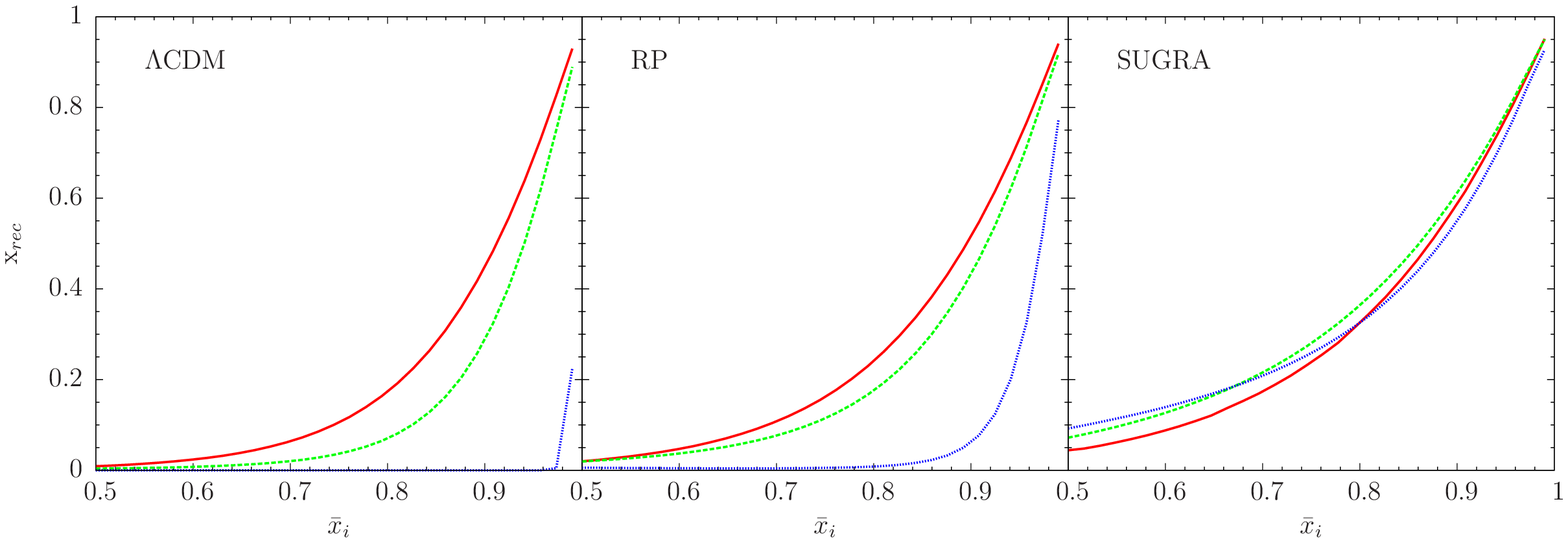}
\caption{The ionized volume fraction.  
In each panel (corresponding to different cosmological models), the
curves refer to the volume fraction contained in recombination-limited
bubbles during the overlap phase at different redshifts: $z=6$ (solid
line), $z=9$ (dashed line) and $z=12$ (dotted line). The results are
computed assuming the MHR00 model with $\lambda_{0}=60$ kms$^{-1}$.}
\label{fig:7d}
\end{figure*}

We can now investigate how the recombination limit due to the IGM
clumpiness affects the geometry and the evolution of the ionized
regions in the quintessence universes.  As already said, in doing
this, we assume that the results of the MHR00 simulations obtained for
a standard cosmology are still valid for a dynamic dark
energy-dominated universe, since we do not expect large differences
for the IGM volume-weighted density distribution.  As discussed above,
the dark energy component causes the matter fluctuations to grow
earlier. Hence, since the recombination rate depends on the inner
overdensity of the bubbles, recombination is strong already at smaller
scales in the quintessence universe, compared to the $\Lambda$CDM
case. Thus, the HII regions reach the equilibrium on scales smaller
than in the standard model. This is clear in Fig.~
\ref{fig:5d}, where we present three different stages of reionization.
While the ionization threshold does not significantly change between
$\Lambda$CDM and SUGRA models, the recombination barrier $\delta_{\rm
rec}$ extends to smaller scales in the quintessence universes. This
effect is more prominent at the late stages of reionization process,
since the bubbles reach the equilibrium on scale of the order of
$20-30$ Mpc in the $\Lambda$CDM universe instead of the $\sim 10$ Mpc
predicted for SUGRA.  In dynamic dark energy universes, the
`earlier' (in term of comoving scales) recombination barrier involves
a smaller value for $R_{\rm max}$, computed as the cross-point of
$\delta_{x}$ and $\delta_{\rm rec}$.  The same trend was already
evident in Fig.~\ref{fig:3d}, where the assumed values for $\zeta$
were such that $\bar{x}_{i}(z=6)=1$.  A peculiarity with respect to
the standard model is the discontinuous recombination barrier found
for the SUGRA cosmology, that crosses the ionization threshold more
than once. To avoid further complications to the model, we choose to
set $R_{\rm max}$ to the mean value of those achieved at the
cross-points (which are in any case very close to each other).

\begin{figure}
\includegraphics[angle=0,height=9cm,width=8cm]{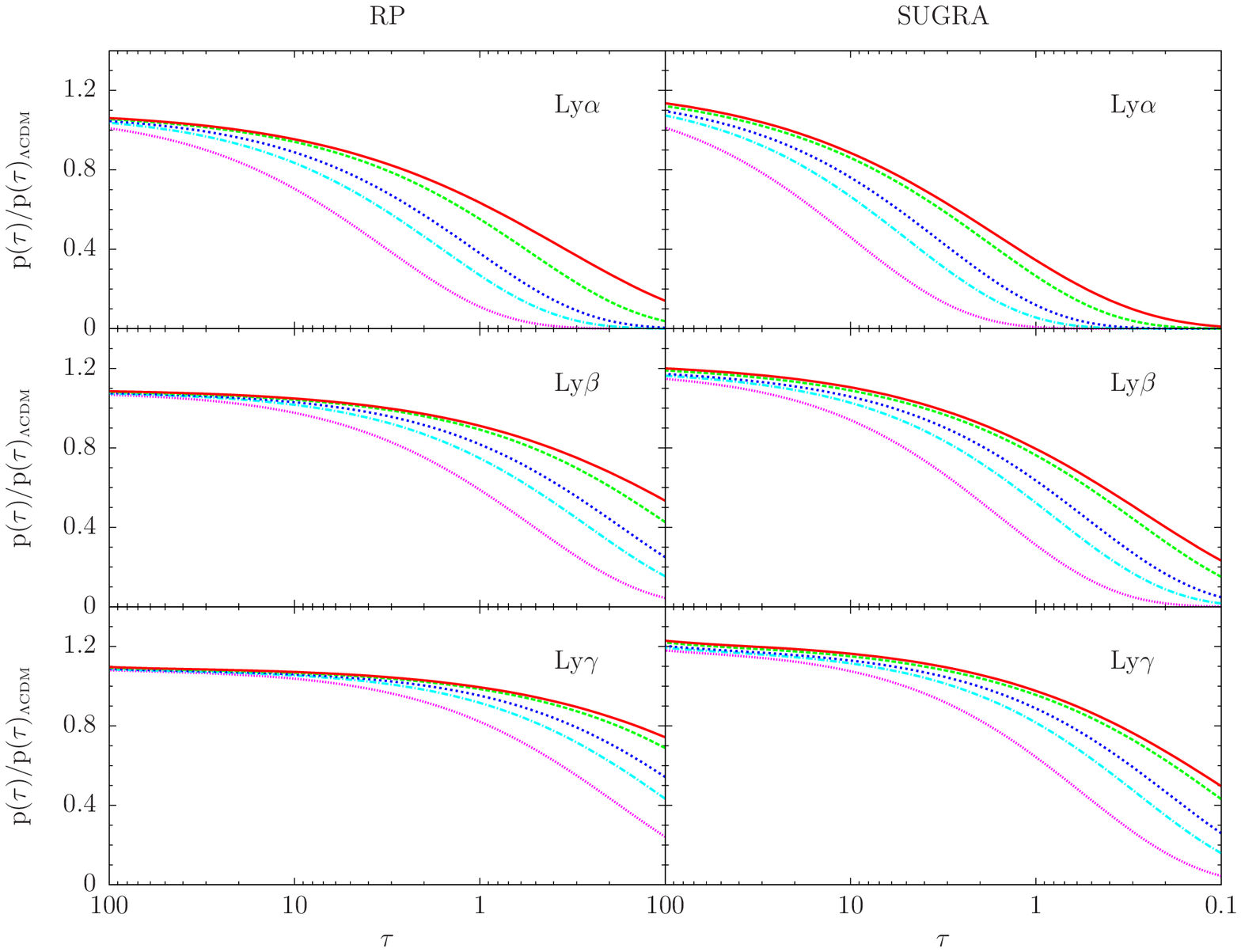}
\caption{
The Lyman transitions.  The probability distribution of the IGM
optical depth $p(\tau)$ for the quintessence models is compared to
that in the `standard' universe ($p(\tau)_{\Lambda CDM}$).  Results
for the RP and SUGRA models are shown in the panels in the left and
right columns, respectively, and refer to Lyman-$\alpha$,
Lyman-$\beta$ and Lyman-$\gamma$, in the different panels, from top to
bottom.  The curves are computed in the case of complete reionization
($\bar{x}_{i}=0.95$) at $z=6$ for different values of $R_{\rm max}$:
10 Mpc (dotted line), 20 Mpc (dotted-dashed line), 30 Mpc
(short-dashed line), 60 Mpc (long-dashed line) and 600 Mpc (solid
line).}
\label{fig:8d}
\end{figure}

Fig.~\ref{fig:6d} shows an important effect on the bubble distribution
of the fact that the maximum radius is smaller, illustrated at $z=8$
for different stages of reionization.  As a result, the HII regions
with $M<M_{\rm max}$ tend to pile up on $\la 10$ Mpc scales in the
SUGRA universe, in particular at the end of reionization. Furthermore,
the drop of the ionization threshold causes an increase of the volume
fraction contained in the recombination-limited regions.

As mentioned before, for a random point in the IGM, the reionization
process can be considered complete when the point joins a sphere with
$M=M_{\rm max}$. Then, since $R_{\rm max}$ is smaller in the
quintessence cosmologies, the density of the points belonging to a
region with $M>M_{\rm max}$ increases and the volume fraction inside
bubbles larger than $M_{\rm max}$ becomes larger. As shown in
Fig.~\ref{fig:7d}, $x_{\rm rec}$ increases moving from the
$\Lambda$CDM to the SUGRA models as reionization proceeds involving
different `epochs of overlapping'. In this case, $x_{\rm rec}\sim 0.5$
is reached earlier in the quintessence models than in the standard
universe.  This effect is analogous to that discussed by F05 for a
small mean free path of the ionizing photons.

As illustrated before, the bubble morphology and the IGM ionization
state affect the IGM optical depth $\tau$ and consequently the
transmission of the Lyman-$\alpha$ flux. To investigate how the IGM
optical depth distribution depends on $R_{\rm max}$ in the
quintessence models, we compute the probability distribution $p(\tau)$
of the IGM optical depth for the Ly-$\alpha,\beta,\gamma$ transmission
following equation (\ref{eq:12c}). In Fig.~\ref{fig:8d} we compare
$p(\tau)$ for the RP and SUGRA scenarios to that obtained considering
the $\Lambda$CDM model, $p(\tau)_{\scriptscriptstyle\rm \Lambda CDM}$,
for different values of $R_{\rm max}$. The transmission is lower for
small values of $\tau$. We find that the trends with $R_{\rm max}$ for
the Lyman-$\beta$ (central panels) and Lyman-$\gamma$ (lower panels)
are analogous to that for the Lyman-$\alpha$ one (upper panels).


\section{Conclusions} \label{sect:conclu}

The purpose of this work is to give a picture of the reionization
epoch in the universes dominated by a dynamic dark energy component at
late epochs, tracing the HII regions evolution using an analytic
approach based on the hierarchic growth of matter fluctuations. In
doing this, we consider two cosmological models in which the dark
energy density varies with time driven by the \citet{peebles2003} and
\citet{brax2000} potentials. Then, we used the analytic
approach proposed by F05 to outline the main
differences between the evolution of bubbles in the quintessence
models and in the standard $\Lambda$CDM cosmology. Our results can be
summarized as follows.

\begin{enumerate}
\renewcommand{\theenumi}{(\arabic{enumi})}

\item The growth of density fluctuations
occurs earlier and the ionization barrier $\delta_{x}$ is lower in the
RP and SUGRA universes compared to the $\Lambda$CDM one.  This causes
a strong increase of the high-density regions with respect to the
$\Lambda$CDM case at the same epoch.

\item Neglecting the recombination limit, the characteristic size of
the HII regions is smaller in the RP and SUGRA cases at the early
stage of reionization, but the difference is weakened as reionization
proceeds.

\item In the recombination limit, the early growth of the matter
fluctuations causes the increase of the IGM clumpiness and the inner
recombination of the bubbles becomes more efficient. As a consequence,
the HII regions reach the ionization equilibrium on slightly smaller
scales and the bubble abundance tends to increase.  The IGM volume
fraction contained in bubbles larger than $R_{\rm max}$ increases
requiring an earlier `epoch of overlap' in the quintessence universe
compared to $\Lambda$CDM.

\item The main effect on the high-$z$ QSO radiation transmission due
to the different evolution of the HII regions is the lower Lyman flux
absorption at small optical depths in RP and SUGRA cosmologies
compared to the $\Lambda$CDM model.

\end{enumerate}

\section*{acknowledgements}

We acknowledge financial contribution from contracts ASI-INAF
I/023/05/0, ASI-INAF I/088/06/0 and INFN PD51.  We thank Steven
Furlanetto, Peng Oh, James Bolton, Enzo Branchini and Micol Bolzonella
for useful discussions. We thank the anonymous referee for
her/his constructive comments.

\newcommand{\noopsort}[1]{}

\label{lastpage}
\end{document}